\documentclass{article}

\usepackage{arxiv}

\usepackage[utf8]{inputenc} 
\usepackage[T1]{fontenc}    
\usepackage{hyperref}       
\usepackage{url}            
\usepackage{booktabs}       
\usepackage{amsfonts}       
\usepackage{nicefrac}       
\usepackage{microtype}      
\usepackage{lipsum}

\usepackage{graphicx}
\usepackage{tabularx}
\usepackage{multirow}
\usepackage{color,array,dcolumn}
\usepackage{amsmath}
\usepackage{bbold}
\usepackage{multirow}
\usepackage{subcaption}

\newcommand{\OP}[1]{\textcolor{black}{#1}}

\newcommand{\Eq}[1]{Eq.~(\ref{#1})}

\newcolumntype{L}[1]{>{\raggedright\let\newline\\\arraybackslash\hspace{0pt}}m{#1}}
\newcolumntype{C}[1]{>{\centering\let\newline\\\arraybackslash\hspace{0pt}}m{#1}}
\newcolumntype{R}[1]{>{\raggedleft\let\newline\\\arraybackslash\hspace{0pt}}m{#1}}

\title{End-to-end  learning for Variational Data Assimilation Models and Solvers}

\author{
  R. Fablet\thanks{Corresponding author} \\
  IMT Atlantique\\ UMR CNRS Lab-STICC,
  Brest, FR \\
  \texttt{ronan.fablet@imt-atlantique.fr} \\
   \And
  B. Chapron \\
  Ifremer\\ UMR CNRS LOPS,
  Brest, FR \\
  \texttt{bertrand.chapron@ifremer.fr} \\
  \And
  L. Drumetz \\
  IMT Atlantique\\ UMR CNRS Lab-STICC,
  Brest, FR \\
  \texttt{lucas.drumetz@imt-atlantique.fr} \\
   \And
  E. Memin \\
  INRIA Rennes\\ UMR CNRS IRMAR,
  Rennes, FR \\
  \texttt{etienne.memin@inria.fr} \\
   \And
  O. Pannekoucke \\
  INPT-ENM\\ UMR CNRS CNRM, CERFACS,
	Toulouse, FR\\
  \texttt{olivier.pannekoucke@meteo.fr} \\
   \And
  F. Rousseau \\
  IMT Atlantique\\ UMR INSERM Latim,
  Brest, FR \\
  \texttt{francois.rousseau@imt-atlantique.fr} \\
}

\begin{document}
\maketitle

\begin{abstract}
This paper addresses variational data assimilation from a learning point of view. Data assimilation aims to reconstruct the time evolution of some state given a series of    observations, possibly noisy and irregularly-sampled. Using automatic differentiation tools embedded in deep learning frameworks, we introduce end-to-end neural network architectures for data assimilation. It comprises two key components: a variational model and a gradient-based solver both implemented as neural networks. A key feature of the proposed end-to-end learning architecture is that we may train the NN models using both supervised and unsupervised strategies.  
Our numerical experiments on Lorenz-63 and Lorenz-96 systems report significant gain w.r.t. a classic gradient-based minimization of the variational cost both in terms of reconstruction performance and optimization complexity. Intriguingly, we also show that the variational models issued from the true Lorenz-63 and Lorenz-96 ODE representations may not lead to the best reconstruction performance. We believe these results may open new research avenues for the specification of assimilation models in geoscience.
\end{abstract}

\keywords{variational data assimilation \and geophysical dynamics \and partial observations \and dynamical model \and end-to-end learning \and meta-learning \and neural networks}

\section{Introduction}
\label{sec:intro}

In geoscience, the reconstruction of the dynamics of a given state or process from a sequence of partial and noisy observations of the system is referred to as a data assimilation issue. Data assimilation is at the core of a wide range of applications, including operational ones, with the aim to make the most of available observation datasets, including for instance both in situ and satellite-derived data, and state-of-the-art numerical models \cite{evensen_data_2009}. Broadly speaking, a vast family of data assimilation methods comes to the minimization of some energy or functional which involves two terms, a dynamical prior and an observation term. We may distinguish two main categories of data assimilation approaches \cite{evensen_data_2009}: variational data assimilation and statistical data assimilation. Variational data assimilation relies on a continuous state-space formulation and results in a gradient-based minimization of the defined variational cost. By contrast, statistical data assimilation schemes generally relies on iterative formulations of stochastic filtering techniques such as Kalman and particle filters. 

Whereas data assimilation naturally applies to model-driven settings, where the formulation of the dynamical and observation models follow from some physical expertise, data-driven strategies have emerged relatively recently \cite{lguensat_analog_2017,bocquet_bayesian_2020} as possible alternatives. The focus has mainly been given to the identification of data-driven representations of the dynamical model, which may be directly plugged into state-of-the-art assimilation frameworks, especially statistical ones \cite{lguensat_analog_2017,ouala_neural-network-based_2018,bocquet_bayesian_2020}. Recent advances have also been reported for the data-driven identification of the dynamical model from partial, irregularly-sampled and/or noisy observation data \cite{nguyen_assimilation-based_2020,bocquet_bayesian_2020,raissi_deep_2018}. 

Here, we aim to further explore how data-driven strategies and associated machine learning schemes may be of interest for data assimilation issues. Especially, end-to-end learning strategies aim to build so-called end-to-end neural network architectures so that one can learn all the components of the architecture w.r.t. some target to be predicted from input data, whereas the traditional approach usually relies on designing each component relatively independently. Many fruitful applications of end-to-end learning strategies have been recently reported in the deep learning literature \cite{schwartz_deepisp_2019,busta_deep_2017,dieleman_end--end_2014}, including for solving inverse problems in image and signal processing. In this work, we introduce a novel end-to-end learning framework for data assimilation, which uses as inputs a sequence of observations and delivers as outputs a reconstructed state sequence.
Following a variational assimilation formulation, our key contributions are as follows: 
\begin{itemize}
    \item the proposed end-to-end learning architecture combines two main components, a first neural-network architecture dedicated to the definition and evaluation of the variational assimilation cost and a second neural-network architecture corresponding to a gradient-based solver of some target loss function. The latter exploits ideas similar to optimizer learning \cite{andrychowicz_learning_2016,vilalta_perspective_2002,li_learning_2018} and directly benefits from automatic differentiation tools embedded in deep learning frameworks, so that we do not need to derive explicitly the adjoint operator of the considered dynamical model;
    \item Given some training criterion, this end-to-end learning architecture provides new means to train all the unknown parameters of the considered neural network architectures. Interestingly, we may consider as training criterion the classic observation-driven data assimilation cost as well as a classic reconstruction error assuming we are provided with groundtruthed data; 
    \item We report numerical experiments on Lorenz-63 and Lorenz-96 dynamics, which support the relevance of the proposed framework w.r.t. classic variational data assimilation schemes.
    Trained NN architectures may both speed up the assimilation process as NN solvers may greatly reduce the number of gradient-based iterations. They may also lead to a very significant improvement of the reconstruction performance if groundtruthed data are available;
    \item Intriguingly, our experiments question the way dynamical priors are defined and calibrated in data assimilation schemes and suggest that approaches based on the sole forecasting performance of the dynamical model may not be optimal for assimilation purposes.  
 \end{itemize}
This paper is organized as follows. Section \ref{sec:pb} introduces the considered variational formulation within a deep learning paradigm. We detail the proposed end-to-end learning framework in Section \ref{sec:learning}. Numerical experiments on Lorenz-63 and Lorenz-96 dynamics are reported in Section \ref{sec:exp}. We further discuss our key contributions in Section \ref{sec:conlusion}.

\section{Problem statement}
\label{sec:pb}

We introduce in this section the considered variational formulation which provides the basis for the proposed end-to-end learning framework. Let us consider the following continuous state-space formulation
\begin{equation}
\label{eq: state space}
\left \{\begin{array}{ccl}
    \displaystyle \frac{\partial x(t)}{\partial t} &=& {\cal{M}}\left (x(t) \right )+ \eta(t)\\~\\
    y(t) &=& x(t) + \epsilon(t), \forall t \in\{t_0,t_0 + \Delta t....,,t_0 + N \Delta t \} \\
\end{array}\right.
\end{equation}
with $x$ the considered time-dependent state, ${\cal{M}}$ the dynamical model and $\{t_0,t_0 + \Delta t....,,t_0 + N \Delta t \}$ the observation times. Observations $\{y(t_i)\}$ may only be partial, meaning that some components of $y(t_i)$ are not observed. Let $\Omega _{t_i}$ denote the domain corresponding to the observed components of $y(t_i)$ at time $t_i$. $\Omega _{t_i}$ may refer to a spatial domain for spatio-temporal dynamics or a list of indices for multivariate state sequence. Processes $\eta$ and $\epsilon$ represent respectively model errors and observation errors.

The data assimilation issue, i.e. the reconstruction of the hidden state sequence $x$ given a series of observations $\{y(t_i)\}$, may be stated as the minimization of the following cost
\begin{equation}
\label{eq: var cost1}
\displaystyle
U_\Phi\left ( x , y , \Omega\right ) = \lambda_1 \sum_i \left \|x(t_i)-y(t_i)\right \|^2_{\Omega _{t_i}} + \lambda_2 \sum_n \left \|x(t_i) - \Phi(x)(t_i) \right \|^2
\end{equation}
where $\| \cdot \|^2_{\Omega}$ stands for the evaluation of the quadratic norm restricted to domain $\Omega$, e.g. $\|u\|^2_\Omega = \int_\Omega u(p)^2dp$ for a scalar 2{\sc d} state $u$. This accounts both for an irregular space-time sampling of the observations as well as an assimilation time step smaller than the observation one. 
$\Phi$ is the flow operator
\begin{equation}
\label{eq: integral flow}
\displaystyle \Phi(x)(t) = x(t-\Delta) + \int_{t-\Delta}^{t} {\cal{M}}\left (x(u)\right) du
\end{equation}
where $\| \cdot \|^2_{\Omega}$ stands for the evaluation of the quadratic norm restricted to domain $\Omega$, e.g. $\|u\|^2_\Omega = \int_\Omega u(p)^2dp$ for a scalar 2{\sc d} state $u$. This accounts both for an irregular space-time sampling of the observations as well as an observation sampling coarser than the assimilation time step.
\OP{In the cost function \Eq{eq: var cost1}, the first term is a weighted measure of the distance from $x$ to the data while the second term measures the distance between the empirical and the theoretical dynamics considering the forecast model as imperfect. This cost function is known as the weak-constraint 4D-Var (see \textit{e.g.} \cite{tremolet_modelerror_2007}). Note here, for the sake of simplicity we do not consider a background term often used to measure the distance from $x$ to a given background state, and corresponding to a Tikhonov regularization.} Using a vectorial formulation, we may rewrite variational cost $U_\Phi  ( x , y , \Omega )$ as:
\begin{equation}
\label{eq: 4DVar discrete}
\displaystyle
U_\Phi \left ( x , y , \Omega\right ) = \lambda_1 \left \|x-y\right \|_\Omega^2 + \lambda_2 \left \|x - \Phi(x) \right \|^2
\end{equation}
with $\|x-y \|_\Omega^2$ the observation term and $\|x - \Phi(x)  \|^2$ the dynamical prior. 
Within a variational assimilation setting, the minimization of this variational cost typically exploits an iterative gradient-based scheme given some initial estimate $x^{(0)}$, the simplest one being a fixed-step gradient descent
\begin{equation}
x^{(k+1)} = x^{(k)} - \alpha \nabla_x U_\Phi \left ( x^{(k)} , y , \Omega\right )
\end{equation}
This requires the computation of gradient operator $\nabla_x U$ using the adjoint method \cite{blayo_reduced_2008,blum_data_2009}. The computational implementation of the adjoint method for operator $\textbf{Id}-\Phi$ may derive from the continuous formulation of $\Phi$ as well as from the discretized version of dynamical model ${\cal{M}}$ to ensure the full consistency between the forward integration of model ${\cal{M}}$ and the adjoint formulation \cite{bannister_review_2017}.

 Within a deep learning framework\footnote{This includes frameworks such as pytorch (\url{https://www.pytorch.org/}),  tensorflow (\url{https://www.tensorflow.org/}) and jax (\url{https://jax.readthedocs.io/}) in Python. We may also cite automatic differentiation tools in Julia (\url{https://www.juliadiff.org/}). Here, we use pytorch and provide our associated code available on \url{https://github.com/CIA-Oceanix/DinAE_VarNN_torchDev}.}, given a neural-network-based (NN) formulation for operator $\Phi$, one may straightforwardly use automatic differentiation tools, typically referred to as the backward step in deep learning, to compute $\nabla_x U$. In this paper, we investigate such NN formulations for operator $\Phi$ and also design NN architectures for the associated solver, i.e. 
an iterative gradient-based inversion algorithm based on variational cost
(\ref{eq: 4DVar discrete}). Interestingly, the resulting end-to-end NN architecture provides means both to learn a solver for a predefined variational setting such as (\ref{eq: 4DVar discrete}) as well as to jointly learn operator $\Phi$ and the solver w.r.t. some reconstruction criterion, which may be different from variational cost (\ref{eq: 4DVar discrete}).

\section{End-to-end learning framework}
\label{sec:learning}

In this section, we detail the proposed end-to-end learning framework based on the NN implementation of variational formulation (\ref{eq: 4DVar discrete}). We first introduce two different types of parameterizations for operator $\Phi$ in (\ref{eq: 4DVar discrete}), which refer to explicit ODE-based and PDE-based representations (Section \ref{ss:ODE operator}) and constrained CNN representations (Section \ref{ss:CNN operator}). We then detail the  NN-based solver for the targeted minimization issue.


\subsection{Explicit ODE/PDE-based formulation for operator $\Phi$}
\label{ss:ODE operator}

Here, we assume that we know the dynamical operator ${\cal{M}}$ in (\ref{eq: state space}), or at least the parametric family it belongs to ${\cal{M}} \in {\cal{F}}=\left \{ {\cal{M}}_\theta \mbox{ with } \theta \in {\cal{A}} \subset {\cal{R}}^N\right \}$ with $N$ the number of parameters of the model:
\begin{equation}
\frac{\partial x(t)}{\partial t} = {\cal{M}}_\theta\left (x(t) \right )
\end{equation}

Considering an explicit Euler integration scheme, operator $\Phi$ is defined as:
\begin{equation}
\Phi(x)(t) = x(t-\Delta) + \Delta \cdot {\cal{M}}_\theta\left (x(t-\Delta) \right )
\end{equation}
with $\Delta$ the integration time step. This resorts to a residual block \cite{he_deep_2016} given the NN-based implementation of model  ${\cal{M}}_\theta$. We may refer the reader to Section \ref{sec:exp} for such examples with Lorenz-63 and Lorenz-96 dynamics.

We may also consider higher-order explicit schemes similarly to \cite{dormand_family_1980}. For example, using a fourth-order Runge-Kutta scheme, operator $\Phi$ is given by:
\begin{equation}
\Phi(x)(t) = x(t-\Delta) + \Delta \cdot \sum_{i=1}^4 \alpha_i k_i \left ( x(t-\Delta), {\cal{M}}_\theta \right )
\end{equation}
where $\alpha_1=\alpha_4=1/6$, $\alpha_2=\alpha_3=2/6$, $k_i \left ( x(t), {\cal{M}}_\theta \right )=  {\cal{M}}_\theta ( x(t) + \beta_i \cdot\Delta\cdot k_{i-1} )$ with $ k_{0}=0$, $\beta_1=\beta_2=\beta_3=1/2$ and $\beta_4=1$. In such formulations, operator $\Phi$ computes a one-step ahead prediction of the input state sequence. We may emphasize that the size of the output of the NN-based implementation of operator $\Phi$ is the same size as the input sequence. In a similar fashion, this also applies to PDE formulations and could be combined to the automatic generation of NN architectures from symbolic PDEs as proposed in \cite{pannekoucke_pde-netgen_2020}.

\subsection{Constrained CNN formulation for operator $\Phi$}
\label{ss:CNN operator}
We also consider CNN architectures for operator $\Phi$. 
CNN architectures whose possible parameterization includes the identity operator $\Phi(x) = x$ shall be excluded as they would result in meaningless priors in minimisation of energy (\ref{eq: 4DVar discrete}). Among classic architectures satisfying this constraint, we may cite auto-encoders such that
\begin{equation}
  \label{eq: constrained CNN}
  \Phi(x) = \phi \left ( \psi(x) \right )
\end{equation}
where operator $\psi$ maps the input variable $x$ to a lower-dimensional space. When considering dynamical systems, auto-encoders may not seem highly relevant to deal with fine-scale processes. Here, following \cite{fablet_end--end_2019}, we focus on an other class of CNN architectures where
$\psi$ is a one-layer CNN where the central values of all convolution kernels is set to zero such that $\psi(x)(s)$ at position $s$ does not depend on variable $x(s)$. Here, $s$ typically refers to a tuple $(t,k)$ of a time index $t$ and a feature index $k$. It would also apply to multivariate space-time tensors. $\phi$ is a CNN which composes a number of convolution and activation layers where the kernel size of all convolution layers is 1 along time and/or space dimensions of process $x$.
We refer to these architectures as GENN (Gibbs Energy Neural Network) as they can be regarded as a NN implementation of Gibbs energies \cite{perez_markov_1998}. 

Intesretingly, within this category of CNN representations for operator $\Phi$, we may consider multi-scale representations to capture patterns of interest at different scales. Here, we consider two-scale representations
\begin{equation}
\label{eq: 2-scale GENN}
    \Phi(x) = Up \left ( \Phi_1 \left ( Dw(x) \right )\right ) + \Phi_2(x)
\end{equation}
where operators $\Phi_{1,2}$ follow the constrained parameterization introduced above. 
$Dw$ is a downsampling operator implemented as an average pooling layer and $Up$ an upsampling operator implemented as a ConvTranspose layer.

\subsection{End-to-end architecture}
\label{ss:NN solver}

\begin{figure}[tbp]
    \centering
    \fbox{\includegraphics[width=13cm]{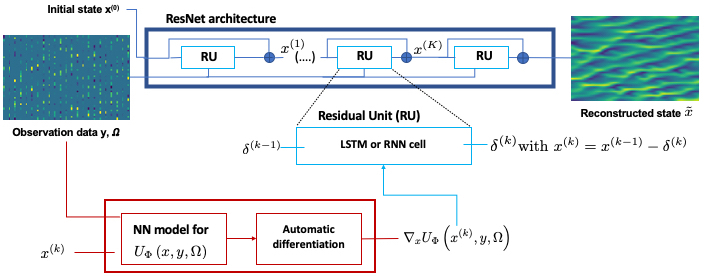}}
    \caption{{\bf Sketch of the proposed end-to-end architecture:} given a partial observation $y$ with missing data mask $\Omega$ and an initialization $x^{(0)}$ for the unknown state $x$ to be reconstructed, the proposed neural network architecture relies on an iterative gradient-based solver. It exploits a residual architecture with a predefined number of residual blocks, where the k$^{th}$ residual unit uses as input the gradient $\nabla_x U_\Phi \left ( x^{(k-1)},y \Omega \right)$ of variational cost $U_\Phi$ w.r.t. state $x$ evaluated  for the output of the previous residual step. Gradient $\nabla_x U_\Phi \left ( x^{(k-1)},y \Omega \right)$ derives from automatic differentiation tools embedded in deep learning frameworks, e.g. autograd function in pytorch.}
    \label{fig: sketch 4DVarNN}
\end{figure}

Given a NN formulation for operator $\Phi$, we design a NN solver for the targeted minimization based on criterion (\ref{eq: 4DVar discrete}). Using automatic differentiation tools embedded in deep learning framework\footnote{Here, we use pytorch. You may refer to the online code available at https://github.com/CIA-Oceanix for details on the use of autograd functions in pytorch.}, we can evaluate the gradient of variational cost (\ref{eq: 4DVar discrete}) w.r.t. variable $x$, denoted as $\nabla_x U$. Using similar ideas as meta-learning schemes \cite{andrychowicz_learning_2016,hospedales_meta-learning_2020}, we can design recurrent neural networks to implement gradient-based solvers for the targeted data assimilation issue. Here, we investigate two types of solvers corresponding to the following iterative updates:
\begin{itemize}
    \item {\bf a LSTM-based solver}, which updates the state at the $k^{th}$ iteration as
\begin{equation}
\label{eq: lstm update}
\left \{\begin{array}{ccl}
     g^{(k+1)}& = &  LSTM \left[ \alpha \cdot \nabla_x U_\Phi \left ( x^{(k)},y , \Omega\right),  h(k) , c(k) \right ]  \\~\\
     x^{(k+1)}& = & x^{(k)} - {\cal{L}}  \left( g^{(k+1)} \right )  \\
\end{array} \right.
\end{equation}
where $\alpha$ is a scalar parameter, $\{h(k) , c(k)\}$  the internal states of the LSTM model and ${\cal{L}}$ a linear layer to map the LSTM output to the space spanned by state $x$. 
    \item {\bf a CNN solver}, which updates the state at the $k^{th}$ iteration as
\begin{equation}
\label{eq: cnn update}
\left \{\begin{array}{ccl}
     g^{(k+1)}& = & {\cal{C}} \left[ \alpha \cdot \nabla_x U_\Phi \left ( x^{(k)},y , \Omega\right),  g^{(k)} \right ]\\~\\
     x^{(k+1)}& = & x^{(k)} - {\cal{A}}  \left( g^{(k+1)} \right )  \\
\end{array} \right.
\end{equation}
where ${\cal{C}}$ is a CNN with a sequence of convolutional layers with Relu activations which uses as inputs the concatenation of gradient  $\nabla_x U_\Phi \left ( x^{(k)},y \Omega \right)$ and previous update $g^{(k)}$. Here, ${\cal{A}}$ refers to an atan activation to avoid exploding updates, especially in the early steps of the learning process.
\end{itemize}
These two types of updates are used as residual blocks to design a residual network (ResNet) with a predefined number of iterations (typically, from 5 to 20 in our experiments). Overall, as sketched in Fig.\ref{fig: sketch 4DVarNN} the resulting end-to-end architecture uses as inputs an initial state $x^{(0)}$, an observation series $y$ and the associated observation domain $\Omega$ to account for missing values and aims to reconstruct the hidden state $x$. In terms of neural network models, it combines a NN model for variational cost $U_\Phi(x,y)$ and a ResNet solver, denoted as $\Gamma$.
Let us denote by $\Psi_{\Phi,\Gamma}(x^{(0)},y,\Omega)$ the resulting end-to-end model.

\subsection{Learning setting}

Given the proposed end-to-end architecture, we may consider different learning strategies. The first strategy
assumes that NN operator $\Phi$ is known. We may be provided with an ODE or PDE representation for dynamical model ${\cal{M}}$ or we may perform as a preliminary step the supervised identification of operator $\Phi$ using some representative dataset for state sequence $x$ \cite{fablet_bilinear_2018,raissi_physics-informed_2019,pannekoucke_pde-netgen_2020}. In such a situation, given some observation dataset comprising a number of observation series $\{y_1,...,y_N\}$ with associated missing data masks $\{\Omega_1,...,\Omega_N\}$, we can address the learning of NN solver $\Gamma$ as the minimization of variational cost (\ref{eq: 4DVar discrete}), which leads to the following learning loss
\begin{equation}
\label{eq: obs loss}
\displaystyle
{\cal{L}} = \sum_n U_\Phi \left ( \Psi_{\Phi,\Gamma} (x^{(0)}_n,y_n,\Omega_n),y_n, \Omega_n \right)
\end{equation}
where parameter $\lambda_{1,2}$ in the definition of energy $U_\Phi$ in (\ref{eq: 4DVar discrete}) are set a priori. This learning strategy is the standard criterion considered in variational data assimilation. It may be regarded as a non-supervised setting in the sense no groundtruthed data is available for the reconstruction of state $x$ from observation data $(y,\Omega)$.

Interestingly, we may also consider a supervised setting where the training dataset comprises observation series $\{y_1,...,y_N\}$, associated missing data masks $\{\Omega_1,...,\Omega_N\}$ and true states $\{x_1,...,x_N\}$. Here, we can consider as learning loss the minimization of the reconstruction error
\begin{equation}
\label{eq: obs loss}
\displaystyle
{\cal{L}} = \sum_n  \left \| x_n - \Psi_{\Phi,\Gamma} \left (x^{(0)}_n,y_n,\Omega_n \right) \right \|^2
\end{equation}
This is a classic supervised strategy where we aim to train the end-to-end architecture so that the reconstruction error of the true state given the observation sequence is minimized. We may point out that, in this supervised setting, the gradient used as input in the trainable solver is not the gradient of the training loss but the gradient of the variational cost, whose computation only involves observation data and not the true states. 

Given these learning losses, we implement all models using pytorch framework and the Adam optimizer. We typically increase incrementally the number of iterations of the gradient-based NN Solver (typically from 5 iterations to 20 ones). We let the reader refer to the code available online\footnote{https://github.com/CIA-Oceanix} for additional details on the implementation.

\section{Numerical experiments}
\label{sec:exp}

This Section reports numerical experiments with the proposed framework for Lorenz-63 and Lorenz-96 systems, which are widely considered for demonstration and evaluation purposes in data assimilation and geoscience which are among the typical case-studies considered in recent data-driven and learning-based studies \cite{lguensat_analog_2017,raissi_deep_2018,bocquet_bayesian_2020}.

\begin{table*}[tb]
    \footnotesize
    \centering
    \begin{tabular}{|C{2.5cm}|C{2.25cm}|C{2cm}|C{1.75cm}|C{1.75cm}|C{1.75cm}|}
    \toprule
    \toprule
     \bf Learning scheme&\bf Model &\bf Optim&\bf R-Score&\bf \bf 4DVar-score&\bf \bf ODE-score\\
    \toprule
    \toprule
     No learning&L63-RK4& FSGD & 3.55 & 7.70e-3 & $<$ 1e-4   \\    
    \toprule
     &L63-RK4& aG-Conv &  3.52 &  1.07e-2 & $<$ 1e-4 \\
     Unsupervised&& aG-LSTM & 4.52 & 1.40e-2 & $<$ 1e-4 \\
     learning&L63-GENN$_1$& aG-Conv & & 29.85&  2.0e-4 \\
     && aG-LSTM & 7.05 & xxxx & 2.0e-4 \\
    \toprule
     &L63-RK4& aG-Conv & 2.82 & 9.24e-2 & {\bf $<$ 1e-4} \\
     Supervisd&& aG-LSTM & 4.00 & 9.36e-2 & {\bf $<$ 1e-4} \\
     learning&L63-GENN$_1$& aG-Conv & {\bf 1.34} & 1.10e-1 & 0.16 \\
     && aG-LSTM & 1.62 & 1.27e-1 & 0.14 \\
    \bottomrule
    \bottomrule
    \end{tabular}
    \caption{{\bf Lorenz-63 experiments:} we report the reconstruction performance of the proposed framework using unsupervised and supervised learning schemes.  We consider two types of representations of the dynamics through operator $\Phi$ (see the main text for details): a NN derived from the known ODE using a RK4 integration scheme (L63-RK4) and a two-scale GENN representation (L63-GENN). 
    Regarding NN solver $\Gamma$, we consider two types of architectures using automatic differentiation to compute the gradient of assimilation cost (\ref{eq: 4DVar discrete}) w.r.t. state $x$: LSTM-based solver (aG-LSTM) and CNN-based ones (aG-CNN). 
    As baseline, referred to as FSGD, we report the performance of a fixed-step gradient descent for assimilation cost (\ref{eq: 4DVar discrete}) with the known ODE model and a fourth-order Runge-Kutta (RK4) integration scheme.  We evaluate three performance metrics: the mean square error of the reconstruction of the true state (R-score), the value of the assimilation cost (\ref{eq: 4DVar discrete}) for the reconstructed state $x$ (4DVar-score) and the mean square error of the one-step-ahead prediction of the consider dynamical prior $\Phi$ (ODE-score).}
    \label{tab:l63}
\end{table*}
\subsection{Lorenz-63 dynamics}

We first perform numerical experiments for the assimilation of Lorenz-63 dynamics from partial and noisy observations. Lorenz-63 system is governed by the following three-dimensional ODE:
 \begin{equation}
\label{eq:lorenz-63}
\left \{\begin{array}{ccl}
\frac{dX_{t,1}}{dt} &=&\sigma \left (X_{t,2}-X_{t,2} \right ) \\
\frac{dX_{t,2}}{dt}&=&\rho X_{t,1}-X_{t,2}-X_{t,1}X_{t,3} \\
\frac{dX_{t,3}}{dt} &=&X_{t,1}X_{t,2}-\beta X_{t,3}
\end{array}\right.
\end{equation}
with the following parameterization: $\sigma =10$, $\rho=28$ and  $\beta=8/3$. It generates chaotic patterns \cite{lorenz_deterministic_1963}. In our experiments, we simulate Lorenz-63 time series with a 0.01 time step using a RK45 integration scheme \cite{dormand_family_1980}. We consider time series with 200 time steps.
The observation data are sampled every 8 time steps for the first component of the Lorenz-63 state only and involve a Gaussian additive noise with a variance of 2. 
The test dataset comprises 2000 sequences and the training data with groundtruthed states contain 10000 sequences. 

Regarding NN operator $\Phi$, we consider two architectures as discussed in Section \ref{sec:learning}:
\begin{itemize}
    \item an ODE-based NN representation based on a 4$^{th}$-order Runge-Kutta integration scheme of the ODE. The ODE operator is implemented as bilinear convolutional layers with a total of 9 parameters as proposed in \cite{fablet_bilinear_2018};
    \item a constrained CNN representation. We consider a two-scale NN representation (\ref{eq: 2-scale GENN}), where operators $\Phi_{1,2}$ involve bilinear convolutional blocks with 30 channels. This architecture involves a total of $\approx$ 15000 parameters.
\end{itemize}
As described in Section \ref{sec:learning}, we consider both LSTM-based and CNN-based versions of ResNet solver $\Gamma$. They involve respectively $\approx$ 3000 and $\approx$ 4000 parameters. 
We let the reader refer to the attached code for implementation details\footnote{A notebook for Lorenz-63 system in repository \url{https://github.com/CIA-Oceanix/DinAE_4DVarNN_torch} provides the implementation of the reported experiments.}.  

\begin{figure*}[tb]
 \centering
 \includegraphics[width=12cm]{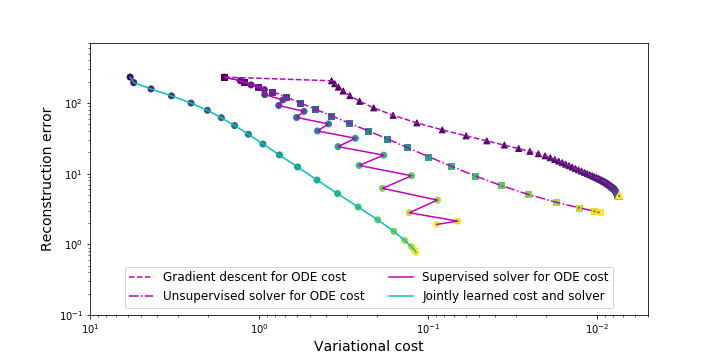}\\
 \includegraphics[width=12cm]{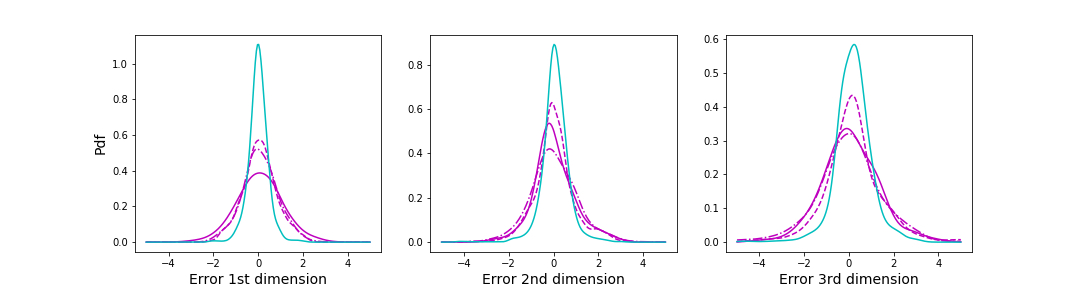} \\  
    \caption{{\bf Reconstruction of Lorenz-63 dynamics.} Upper panel: solvers' energy pathways using different parameterizations for operator $\Phi$ in  (\ref{eq: 4DVar discrete}). We depict along the sequence of iterations of different gradient-based solvers the evolution of the reconstruction error and of the variational cost. For the known Lorenz-63 ODE model, we report these energy pathways for fixed-step gradient descent of variational cost (\ref{eq: 4DVar discrete}) (magenta dashed) and learned NN solvers using  unsupervised (magenta, dashed) and supervised (magenta, dashed-dotted) learning schemes. We also report the energy pathway for a joint supervised learning of a GENN-based parameterization of operator $\Phi$ and of the associated NN solver. Lower panel: associated pdfs of the reconstruction error for each component of the Lorenz-63 state.}  
    \label{fig:res L63a}
\end{figure*}

We report the performance metrics in Tab.\ref{tab:l63} for the different parameterizations of operator $\Phi$ and the associated solvers using either a supervised learning strategy or an unsupervised one (See Section \ref{sec:learning}). As baseline, we  consider a fixed-step gradient descent (FSGD) of assimilation cost (\ref{eq: 4DVar discrete}) using a fourth-order Runge-Kutta integration scheme for the known ODE model. This baseline is also implemented in Pytorch. All trained solvers involve 20 gradient-based iterations. As evaluation metrics, we consider the reconstruction error of the true state (mean square error) (R-score), the assimilation cost (4DVar-score) and the mean square error for the one-step-ahead dynamical prior $\|x-\Phi(x)\|^2$ (ODE-score). To compute comparable assimilation costs across methods, we 
report the assimilation cost for all schemes with $\lambda_1 = 0.01$ and $\lambda_2=1.0$, although these parameters may differ from the values 
learned during the training phase for supervised settings.

From Tab.\ref{tab:l63}, we may first notice that all schemes, which aim at minimizing assimilation cost (\ref{eq: 4DVar discrete}, {\em i.e.} the FSGD and unsupervised solvers, lead to worse reconstruction performance (R-score about 3.5) compared with supervised schemes (R-score up to 1.62). Conversely, the former leads to significantly better 4DVar-scores. These results emphasize that the assimilation cost may not be a very relevant indicator of the reconstruction performance. For instance, the trained aG-Conv solver leads to a slightly better performance than the FSGD (R-score of 3.52 vs. 3.55) but with a greater assimilation cost (4DVar-score of 1.05e-2 vs. 7.70e-3). In the unsupervised setting, the use of a GENN prior results in very poor reconstruction performance, which may relate to a worse prediction performance (ODE-score of 2.0e-4 vs. 3.7e-5 for a fourth-order Runge-Kutta scheme with the true ODE model).

\begin{figure*}[bt]
\begin{center}
\begin{tabular}{cccc}
{\bf Example \#1}&
{\bf Example \#2}&
{\bf Example \#3}&
{\bf Example \#4}\\
\includegraphics[trim={0 40 0 40},clip,width=3.5cm]{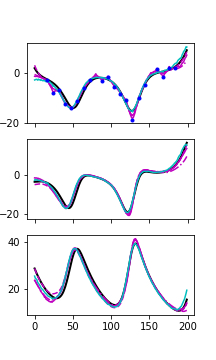}&
\includegraphics[trim={0 40 0 40},clip,width=3.5cm]{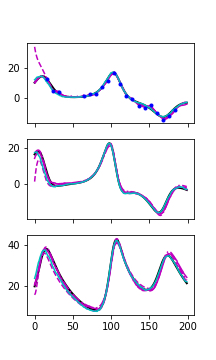}&
\includegraphics[trim={0 40 0 40},clip,width=3.5cm]{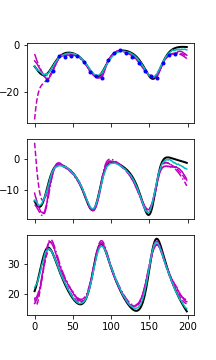}&
\includegraphics[trim={0 40 0 40},clip,width=3.5cm]{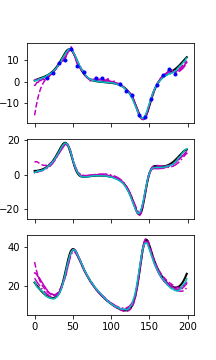}\\
\end{tabular}
\end{center}
    \caption{{\bf Reconstruction examples for Lorenz-63 dynamics}: from left to right, we report four examples of 200-step Lorenz-63 sequences with true states (black, solid), observed values for the first component (blue dots), assimilated states with different schemes using the true ODE model (magenta) and a jointly learned GENN-based operator $\Phi$ and solver (cyan). We let the reader refer to Fig.\ref{fig:res L63a} for the details of the different ODE-based schemes. Each row depicts one of the three components of the Lorenz-63 states.}
    \label{fig:res L63b}
\end{figure*}

Importantly, the supervised schemes greatly  the reconstruction error compared with the baseline (R-score of 1.34 for the best supervised scheme vs. 3.55 for the FSGD scheme). The best performance is indeed reported for a joint learning of operator $\Phi$ and of the associated solver, using a GENN parameterization for operator $\Phi$. The improvement is also significant compared to the supervised learning of a solver using the discretized version of the true ODE model for operator $\Phi$. Intriguingly, the learnt GENN-based operator $\Phi$ is characterized by a relatively poor one-step-ahead prediction error (ODE-score of 0.16 vs. 3.7e-5 for the ODE-based schemes). These results suggest that the best dynamical prior for assimilation purposes might not be the discretization of the true ODE model. The detailed analysis of the values of the assimilation cost further points out that its direct minimization may not be the best strategy to optimize the reconstruction performance.  

We further illustrate these results in Fig.\ref{fig:res L63a} and Fig;\ref{fig:res L63b}. For a randomly-sampled subset of the test dataset, we report in Fig.\ref{fig:res L63a} the energy pathways of different solvers: namely the baseline (FSGD), the best unsupervised and supervised solvers using the true ODE model for operator $\Phi$, and the best supervised solver for a GENN-based operator $\Phi$. Here, the FSGD scheme involves more than 150000 steps, whereas all the learned solvers involve only 20 gradient-based iterations. In these experiments, the CNN-based solver always  reach a better performance than the LSTM-solver. Interestingly, the joint learning of operator $\Phi$ and of the associated solver leads a smooth descent, which may indicate that the training phase converges towards a new variational cost with a better agreement between the minimization of this cost and the reconstruction performance. 

The distributions of the error of the reconstruction for the three components of the Lorenz-63 state in Fig.Fig.\ref{fig:res L63a} along with different examples in Fig.\ref{fig:res L63b} further emphasize the better reconstruction performance of the joint learning of a GENN-based operator $\Phi$ and a aG-Conv solver, which leads to reconstruction patterns very similar to the true ones. 

\subsection{Lorenz-96 dynamics}
The second experiment involves Lorenz-96 dynamics which are governed by the following ODE
\begin{equation}
\label{eq:lorenz-96}
\frac{dx_{t,i}}{dt} = \left (x_{t,i+1}-x_{t,i-2} \right ) x_{t,i-1} - x_{t,i} + F
\end{equation}
with $i$ the index from 1 to 40 and $F$ a scalar parameter set to 8. The above equation involves a periodic boundary constraint along the multivariate dimension of state $x(t)$. In our experiments, we simulate Lorenz-96 time series with a 0.05 time step using a RK45 integration scheme \cite{dormand_family_1980}. We consider time series with 200 time steps. The observation data are sampled every 4 time steps and involve a Gaussian additive noise with a variance of 2. Only 20 of the 40 components of the states are observed according to a random sampling. The test dataset comprises 256 sequences and the training data with groundtruthed states involve 2000 sequences.

Regarding NN operator $\Phi$, we consider two architectures as discussed in Section \ref{sec:learning}:
\begin{itemize}
    \item an ODE-based NN representation, referred to as L96-RK4 based on a 4$^{th}$-order Runge-Kutta integration scheme of the ODE. Our implementation involves bilinear convolutional layers with a periodic boundary conditions. This representation comprises a total of 9 parameters;
    \item a constrained CNN representation referred to as L96-GENN. We consider a two-scale NN representation (\ref{eq: 2-scale GENN}), where operators $\Phi_{1,2}$ involve bilinear convolutional architectures with a total of $\approx$ 50000 parameters.
\end{itemize}
As described in Section \ref{sec:learning}, we consider both LSTM-based and CNN-based versions of solver $\Gamma$. They involve respectively $\approx$ 1000 and X$\approx$ 1500 parameters. We let the reader refer to the code made available for implementation details\footnote{A colab notebook for Lorenz-96 system in repository \url{https://github.com/CIA-Oceanix/DinAE_4DVarNN_torch} describes the implementation of the reported experiments.} 

\begin{table*}[bt]
    \footnotesize
    \centering
    \begin{tabular}{|C{2.5cm}|C{2.25cm}|C{2cm}|C{1.75cm}|C{1.75cm}|C{1.75cm}|}
    \toprule
    \toprule
     \bf Learning scheme&\bf Model &\bf Optim&\bf R-Score&\bf \bf 4DVar-score&\bf \bf ODE-score\\
    \toprule
    \toprule
     No learning&L96-RK4& 4DVar & 1.06 & 1.47e-2 & $<$ 1e-4  \\    
    \toprule
     Unsupervised&L96-RK4& aG-Conv & 1.00  & 1.66e-2 & $<$ 1e-4 \\
     learning&& aG-LSTM & 1.32  & 1.91e-2 & $<$ 1e-4 \\
    \toprule
     &L96-RK4&aG-Conv& 0.82 & 2.12e-2 & {\bf < 1e-4}   \\
     Supervised&&aG-LSTM & 0.97 & 3.15e-2 & {\bf < 1e-4}   \\
     learning&L96-GENN&aG-Conv & 0.49& 7.47e-2 & 12.20e-2  \\
     &&aG-LSTM & {\bf 0.38} & 5.30e-2 & 7.20e-2  \\
    \bottomrule
    \bottomrule
    \end{tabular}
    \caption{{\bf Lorenz-96 experiments:} we report the reconstruction performance of the proposed framework using unsupervised and supervised learning schemes. We let the reader refer to Tab.\ref{tab:l63} and the main text for details on the considered schemes and evaluation criteria.}
    \label{tab:l96}
\end{table*}

\begin{figure*}[tb]
\begin{subfigure}{0.3\textwidth}
\begin{tabular}{c||cc}
    {\footnotesize {\bf True and observed states}} \\
    \includegraphics[trim={55 110 150 110},clip, width=4cm]{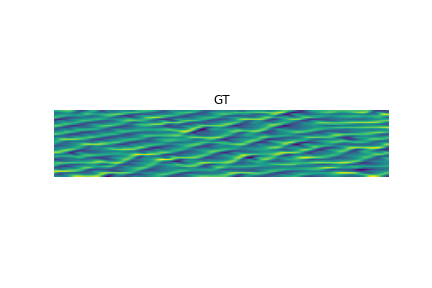}\\
    \includegraphics[trim={55 110 150 110},clip, width=4cm]{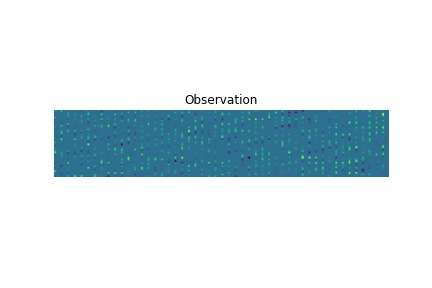}\\
\end{tabular}
\end{subfigure}
\begin{subfigure}{0.7\textwidth}
\includegraphics[height=4.cm]{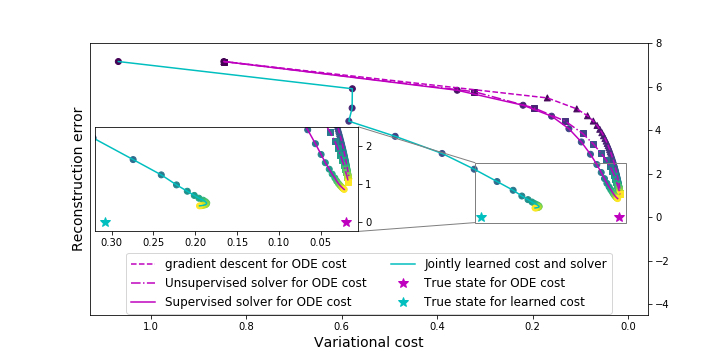}
\includegraphics[height=4.cm]{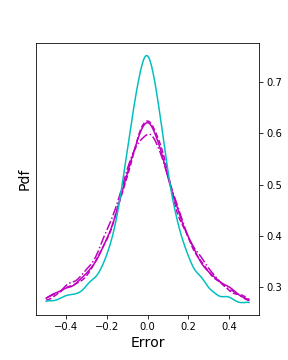}
\end{subfigure}\\
\begin{center}
\begin{tabular}{cccc}
\hline
    ~\\
    \multicolumn{4}{c}{{\footnotesize {\bf Reconstruction examples and associated error maps}}} \\~\\
    \includegraphics[trim={55 110 150 110},clip, width=3.5cm]{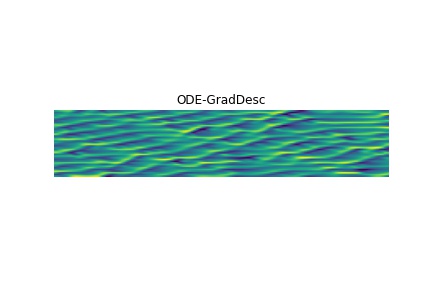}&
    \includegraphics[trim={55 110 150 110},clip, width=3.5cm]{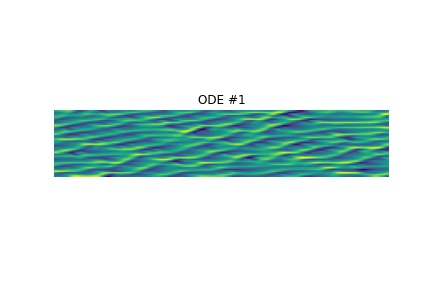}&
    \includegraphics[trim={55 110 150 110},clip, width=3.5cm]{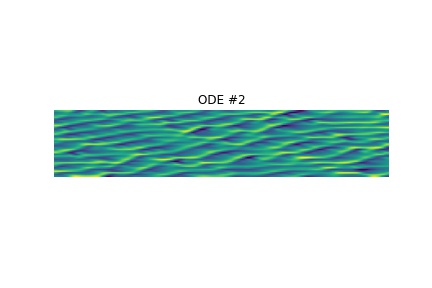}&
    \includegraphics[trim={55 110 150 110},clip, width=3.5cm]{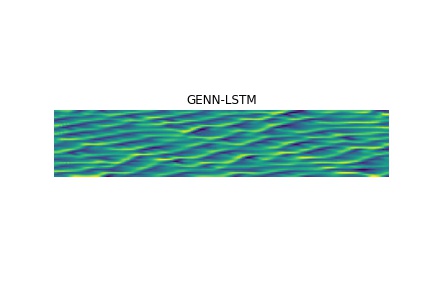}\\
    \includegraphics[trim={55 110 150 110},clip, width=3.5cm]{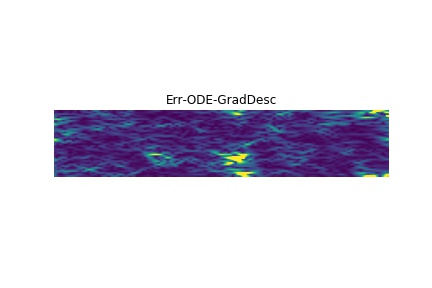}&
    \includegraphics[trim={55 110 150 110},clip, width=3.5cm]{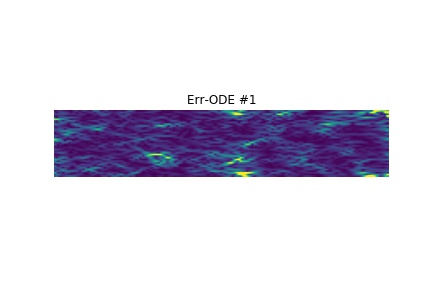}&
    \includegraphics[trim={55 110 150 110},clip, width=3.5cm]{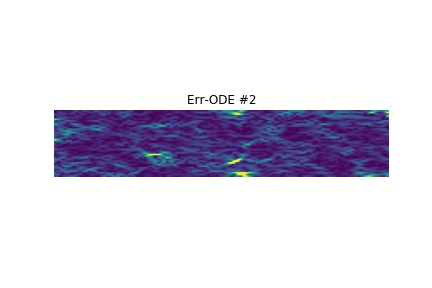}&
    \includegraphics[trim={55 110 150 110},clip, width=3.5cm]{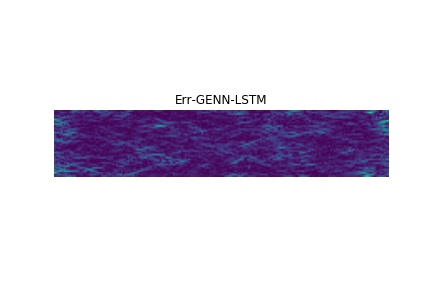}\\
    \footnotesize Gradient descent & \footnotesize Unsupervised solver & \footnotesize Supervised solver & \footnotesize Jointly learned\\
    \footnotesize for ODE cost & \footnotesize for ODE cost & \footnotesize for ODE cost & \footnotesize cost and solver \\
 \end{tabular}
\end{center}
    \caption{{\bf Reconstruction of Lorenz-96 dynamics.} Upper left panel: example of true state sequence and associated observations; Upper center panel: solvers' energy pathways using Lorenz-96 ODE cost or learnable variational cost; Upper right panel: associated pdfs of the reconstruction errors.  Lower panel: We depict the first half of an example of 200-time-step series of 40-dimensional Lorenz-96 states, the x-axis being the time axis; first row, reconstructed state sequence for the unsupervised and supervised solvers for the ODE cost (\ref{eq: 4DVar discrete}) as well as for the jointly learned cost and solver; second row, absolute error maps of each of the four reconstructions. All states and errors are displayed with the same colormap.}
    \label{fig:res L96}
\end{figure*}

We first report in Tab.\ref{tab:l96} a quantitative comparison of different parameterizations of operator $\Phi$ and solver $\Gamma$.  Similarly to Lorenz-63 experiments, the best unsupervised solver for an ODE-based cost reaches the same  reconstruction  performance as the baseline using only 20 gradient-based iterations, to be compared with several thousands for the FSGD solver. Again, the CNN-based solver outperforms the LSTM-based one in the unsupervised setting. We do not report the performance for the GENN-based representation in the unsupervised setting as it behaves very poorly similarly to Lorenz-63 experiments. Besides, we report a very significant improvement in the supervised setting, especially when we jointly learn operator $\Phi$ and solver $\Gamma$ with a relative gain greater than 50\% compared with the baseline. In this case, the best performance is achieved with a LSTM-based solver. Again, the best reconstruction performance does not come with a good one-step-ahead prediction score for the learned GENN-based representation. 

We further illustrate these experiments in Fig.\ref{fig:res L96}. Here, all solvers' energy pathways are consistent with minimization patterns both for the reconstruction error and the assimilation cost. These pathways are very similar for the three solvers using an ODE-based cost. Visually, they lead to similar error distributions. For the example reported in the bottom panel of Fig.\ref{fig:res L96}, they involve similar error patterns though the FSGD solver involves greater  errors. In this respect, the reconstruction issued from the joint learning of a GENN-based operator $\Phi$ and a LSTM-based solver is associated with lower error values.

\section{Related work}
\label{sec:rw}

In this section, we further discuss how the proposed framework relates to previous works according to three different aspects: end-to-end learning framework for the resolution of inverse problems, learning schemes for gradient-based solvers and representation learning for geophysical dynamics.

{\bf End-to-end learning for inverse problems:} A variety of end-to-end architectures have been proposed for solving inverse problems in signal and image processing, for instance for denoising, deconvolution or super-resolution issues \cite{chen_learning_2015,lucas_using_2018,mccann_convolutional_2017,xie_image_2012}. In these studies, proposed schemes generally rely on a global convolutional architecture, but do not explicitly state on one hand a neural-network representation of an energy-based setting and, on the other hand, the gradient-based solver of the consisdered variational formulation. For instance, in \cite{chen_learning_2015}, the end-to-end architecture is inspired by reaction-diffusion PDEs derived from the Euler-Lagrange equation of a variational formulation, but does not explicitly embed such a variational formulation. Conversely, deep learning frameworks and embedded automatic differentiation tools have been considered to solve for inverse problems through the minimization of variational models, where the prior can be given by a pre-trained NN representation \cite{mccann_convolutional_2017}. An other strategy explored in the literature relies on the definition of two independent networks, one corresponding to the generative model and the other one to the inverse model. Raissi et al. \cite{raissi_deep_2018} were among the first to explore such end-to-end architectures for the data-driven identification of ODE/PDE representations. This is similar in spirit to auto-encoder architectures: the decoder aims to solve the inverse of the encoder but it does not explicitly depend on the representation chosen for the encoder. Here, the considered end-to-end architecture exploits the automatic differentiation to explicitly relate the gradient-based inversion model to the neural-network representation of the dynamics. This is regarded as of great interest to improve the overall interpretability of the architecture. Besides, our experiments support that jointly learning  the representation of the dynamics and the associated solver can lead to a very significant improvement compared with pre-training the dynamical prior and solving the resulting variational minimisation.
   
{\bf Learning gradient-based solvers for inverse problems:} A key component of the proposed end-to-end architecture is the gradient-based NN solver for the targeted minimization. This is similar to meta-learning and learning-to-learn strategies \cite{andrychowicz_learning_2016,hospedales_meta-learning_2020}, where one can learn jointly some targeted representation and the optimizer of the training loss. The LSTM-based architecture of the considered gradient-based NN optimizer is similar to the one considered in learning-to-learn schemes. We may notice that we apply here automatic differentiation to compute the gradient w.r.t. the hidden state and not the parameters of the NN representations. 
Besides, in the supervised learning case, the gradient computed using automatic differentiation is not the gradient of the loss to be minimized during the training process but the gradient of assimilation criterion (\ref{eq: 4DVar discrete}). We indeed expect the latter to be a good proxy to minimize the reconstruction error of the true states. The reported numerical experiments support this assumption and suggest extensions to other training losses which could benefit from additional data, not provided as inputs for the computation of variational cost  (\ref{eq: 4DVar discrete}). \OP{Note that in general, the 4DVar cost-function can present multiple local and global minima, except in the particular case where the dynamics is linear (as well as the observational operator that map the state space to the observational space). In order to avoid falling  into local minima or to select the absolute minimum a quasi-static approach can be considered \cite{pires_extending_1996}, based on a successive small increments of the assimilation period. While this issue is not addressed in the present work, a similar quasi-static approach could be considered in the design of the learning gradient-based solver.}

{\bf Representation learning for geophysical dynamics:} As stated in the introduction, the data-driven identification of representations of geophysical dynamics is very active research area. Numerous recent studies have investigated different machine learning schemes for the identification of governing equations of geophysical processes, including sparse regression techniques \cite{brunton_discovering_2016}, neural ODE and PDE schemes \cite{chen_neural_2018,raissi_physics-informed_2019,pannekoucke_pde-netgen_2020} as well as analog methods \cite{lguensat_analog_2017}. Interestingly, advances have been proposed for the data-driven identification of such representations, when the processes of interest are not fully observed, which covers noisy and irregularly-sampled observations, partially-observed system. Reduced-order modeling, which aims to identify lower-dimensional governing equations, also involves very similar studies \cite{champion_data-driven_2019}. These previous works mainly rely on the identification of an ordinary or partial differential equation as this mathematical representation is a very generic one for modeling geophysical dynamics. The proposed end-to-end framework explicitly embeds a representation of the considered geophysical dynamics. In line with previous works cited above, we may consider neural ODE/PDE representations \cite{pannekoucke_pde-netgen_2020,chen_neural_2018,fablet_bilinear_2018}. As we rely on an energy-based setting, we can explore other types of representation. In the reported experiments, we have shown that a multi-scale energy-based representation for the dynamical prior leads to significantly better reconstruction performance than an ODE-based representation at the expense of the relatively coarse approximation of the dynamics. These experiments suggest that, for given process, different representations might be relevant depending on targeted application, especially simulation vs. assimilation. It also raises the question as to whether the representation should be adapted to observation operator. 

{\bf Data assimilation for geophysical flow forecasting:} This work can be adpated to the different classical formulations considered used for data assimilation issues in meteorology or oceanography. Compared with (\ref{eq: 4DVar discrete}), the data assimilation cost usually involves a discrepancy term on the initial condition, called the background term. This term is usually expressed as a Mahalanobis distance and involves the inverse of the background covariance matrix. This matrix plays the role of a smoothing prior on the solution and its specification is essential, yet difficult to fix/parameterize on physical grounds. In the proposed framework, this would correspond to parameterizing and learning this discrepancy term with respect to the initial condition, while the dynamics is fixed and given. While the considered formulation (\ref{eq: 4DVar discrete}) relates to the weak-constraint data assimilation, which accounts for modelling erors, we could also explore the proposed framework for the so-called strong constraint data assimilation, which amounts to stating operator $\Phi$ as the forecasting of the state sequence $\{x(t_i)\}_i$ over the entire time interval from the initial condition $x(t_0)$. Within a weak-constraint data assimilation with a known physical model, the proposed framework may provide new means to learn of a correction error term in the dynamics from a parameterization of operator $\Phi$ as $\Phi
^*+ d\Phi$ with $\Phi^*$ defined as in the strong constraint case and $d\Phi$ as in the reported experiments.


{\bf Learning for computational fluid dynamics:} In the present study, one of the considered NN architecture for operator $\Phi$ derives from a general time evolution model interpreted in terms of iterative integration schemes (i.e. Runge-Kutta, Euler etc.). In computational flow dynamics, splitting methods are usually introduced to deal with incompressiblity and the computation of the pressure forcing term \cite{issa_computation_1986,patankar_numerical_2018}. The introduction of such splitting in the NN architecture should enable us to enforce incompressibility of the solution and thus to strengthen the physical relevance of the solutions. The synergy between NN architectures and numerical schemes is a natural way to enforce some numerical constraints imposed by the deep physical features of the dynamics at stake. For instance, splitting in terms of slow 3D internal baroclinic and fast 2D (depth averaged) barotropic motions, as it is usually done in numerical ocean models, should strongly enforce a natural physical relevance for the learning of ocean dynamics.

\section{Conclusion}
\label{sec:conlusion}

We have introduced an end-to-end learning framework for variational assimilation problems. Assuming the observation operator is known, it combines a neural-network representation of the dynamics and a neural-network solver for the considered variational cost. We may derived the latter from the known differential operator for the considered dynamics. Here, for Lorenz-63 and Lorenz-96 dynamics, we considered NN representations which implement fourth-order Runge-Kutta integration schemes. A similar approach can be considered for PDEs, including with automatic NN generation tools from symbolic PDE expressions as proposed in \cite{pannekoucke_pde-netgen_2020}. Depending on the learning setting, we may learn jointly the two NN representations or only the NN solver assuming the NN representation of the dynamics has been calibrated previously. We provide a pytorch implementation of the proposed framework.

We report numerical experiments which support the relevance of the proposed framework. For Lorenz-63 and Lorenz-96 dynamics, we have shown that we may learn fast gradient-based solver that can reach state-of-the-art performance with only a few gradient iterations (20 in the reported experiments). Interestingly, our findings suggest that, when groundtruthed datasets are available, the joint learning of the NN representation of the dynamics and of the associated NN solver may lead to a very significant improvement of the reconstruction performance. Illustrated here for Lorenz systems, future work shall further explore whether these findings generalize to other systems, especially higher-dimensional ones. 

Our findings also question the design and selection of the dynamical prior in variational assimilation system. They suggest that numerical ODE-based representations optimal in terms of forecasting performance may not be optimal for reconstruction purposes. Future work shall further explore this question on the synergy between the representation of the dynamics and the gradient-based solver to retrieve the best possible state estimate.  The extension of the proposed framework to stochastic representations is also of great interest and may for instance benefit from the definition of  variational functionals from log-likelihood functions where the different terms or parameters may be represented as neural networks.  

\section*{Acknowledgements}
This work was supported by LEFE program (LEFE MANU project IA-OAC), CNES (grant OSTST-MANATEE) and ANR Projects Melody and OceaniX. It benefited from HPC and GPU resources from Azure (Microsoft EU Ocean awards) and from GENCI-IDRIS (Grant 2020-101030).

\bibliographystyle{plain}  

\end{document}